\documentclass[aps,prl,twocolumn,preprintnumbers,superscriptaddress,dblfloatfix,nofootinbib]{revtex4-1}
\usepackage[utf8]{inputenc}
\usepackage{lmodern}
\usepackage{amsmath,amssymb,mhchem}
\usepackage{graphicx}
\usepackage{url}
\usepackage{color}
\usepackage{enumitem}
\usepackage{subfigure}
\usepackage[dvipsnames]{xcolor}
\usepackage[colorlinks=true,breaklinks=true]{hyperref}
\hypersetup{allcolors=[rgb]{0.0 0.0 0.6},linkcolor=[rgb]{0.75 0.05 0.05}}
\usepackage{bm}
\usepackage{epsfig}
\usepackage{amsmath}
\usepackage{amssymb}
\usepackage{slashed}
\usepackage{color}
\usepackage{accents}
\usepackage{ulem}

\usepackage[dvipsnames]{xcolor}
\usepackage[colorlinks=true,breaklinks=true]{hyperref}
\hypersetup{allcolors=[rgb]{0.0 0.0 0.6},linkcolor=[rgb]{0.75 0.05 0.05}}
\usepackage{multirow}

\newcommand{\exclude}[1]{}

\begin{document}

\preprint{IPMU20-0087} 

\title{Test for the Origin of Solar Mass Black Holes}

\author{Volodymyr Takhistov}
\email{vtakhist@physics.ucla.edu}
\affiliation{Department of Physics and Astronomy, University of California, Los Angeles \\ Los Angeles, California, 90095-1547, USA}
\affiliation{Kavli Institute for the Physics and Mathematics of the Universe (WPI), UTIAS \\The University of Tokyo, Kashiwa, Chiba 277-8583, Japan}

\author{George M. Fuller}
\email{gfuller@ucsd.edu}
\affiliation{Department of Physics, University of California, San Diego \\ La Jolla, California 92093-0319, USA
}
\affiliation{Center for Astrophysics and Space Sciences, University of California, San Diego\\ La Jolla, California 92093-0424, USA}

\author{Alexander Kusenko}
\affiliation{Department of Physics and Astronomy, University of California, Los Angeles \\ Los Angeles, California, 90095-1547, USA}
\affiliation{Kavli Institute for the Physics and Mathematics of the Universe (WPI), UTIAS \\The University of Tokyo, Kashiwa, Chiba 277-8583, Japan}
 
\date{\today}

\begin{abstract}
Solar-mass black holes with masses in the range of $\sim 1-2.5 M_{\odot}$ are not expected from conventional stellar evolution, but can be produced naturally via neutron star (NS) implosions induced by capture of small primordial black holes (PBHs) or from accumulation of some varieties of particle dark matter. We argue that a unique signature of such ``transmuted'' solar-mass BHs is that their mass distribution would follow that of the NSs. This would be distinct from the mass function of black holes in the solar-mass range predicted either by conventional stellar evolution or early Universe PBH production. We propose that analysis of the solar-mass BH population mass distribution in a narrow mass window of $\sim 1-2.5\,{\rm M}_\odot$ can provide a simple yet powerful test of the origin of these BHs.
Recent LIGO/VIRGO gravitational wave (GW) observations of the binary merger events
GW190425 and GW190814 are consistent with a BH mass in the range $\sim 1.5-2.6~M_{\odot}$. Though these results have fueled speculation
on dark matter-transmuted solar-mass BHs, we demonstrate that it is unlikely that the origin of these particular events stems from NS implosions. Data from upcoming GW observations will be able to distinguish between solar-mass BHs and NSs with high confidence. This capability will facilitate and enhance the efficacy of our proposed test.
\end{abstract}

\maketitle


The discovery of gravitational waves (GWs)~\cite{Abbott:2016blz} has opened a new window for exploration of black hole (BH) properties, in particular the distribution of their masses. In this Letter we focus on the BH mass parameter space region around a solar mass, where BHs are not expected to be produced by conventional stellar evolution.  However,
primordial black holes (PBHs), produced in the early universe~\cite{Zeldovich:1967,Hawking:1971ei,Carr:1974nx,Khlopov:1985jw,Yokoyama:1995ex,GarciaBellido:1996qt,Kawasaki:1997ju,Khlopov:2008qy,Frampton:2010sw,Kawasaki:2012kn,Kawasaki:2012gk,Kawasaki:2016pql,Carr:2016drx,Inomata:2016rbd,Pi:2017gih,Cotner:2016dhw,Cotner:2016cvr,Deng:2016vzb,Deng:2017uwc,Cotner:2017tir,Inomata:2017okj,Garcia-Bellido:2017aan,Georg:2017mqk,Inomata:2017vxo,Kocsis:2017yty,Ando:2017veq,Cotner:2018vug,Sasaki:2018dmp,Carr:2018rid,Cotner:2019ykd,Kusenko:2020pcg}, could be of solar-mass scale. Furthermore, PBHs with much smaller masses can be captured in a neutron star (NS) and can convert the NS to a solar-mass black hole~(e.g.~\cite{Fuller:2017uyd,Capela:2013yf,Takhistov:2017bpt,Bramante:2017ulk,Takhistov:2017nmt,Capela:2013yf,Genolini:2020ejw}).  Likewise, accumulation of dark matter (DM) inside a neutron star can cause the NS to implode and form a solar-mass BH (e.g.~\cite{Bramante:2017ulk,Kouvaris:2018wnh}).  Here we suggest a powerful yet simple test to distinguish between these possibilities and hence shed light on the origin of solar-mass BHs and DM.

LIGO/Virgo has reported events  GW190425~\cite{Abbott:2020uma} and GW190814~\cite{Abbott:2020khf} suggestive of possible BHs with $\sim 1.5-2.6 M_{\odot}$ masses. While such binary constituents could be NSs~(e.g.~\cite{Godzieba:2020tjn,Tsokaros:2020hli}), they are consistent with solar-mass BHs.  Without identification of clear electromagnetic (EM) signal counterparts or good sensitivity to higher order tidal deformability GW effects, distinguishing between a solar-mass BH and a NS is difficult. Even with observation of an EM counterpart, distinguishing a NS-NS binary merger event like GW170817~\cite{TheLIGOScientific:2017qsa} from a NS-BH event is nontrivial. Improved discrimination between the two can be achieved with accumulation of observed event statistics~\cite{Yang:2017gfb,Chen:2019aiw}.

Astrophysical BHs and NSs are thought to form via gravitational collapse of massive stars in conventional stellar evolution, in particular, as remnants of core-collapse supernovae (SN) \cite{Fryer:2012}. Following core-collapse, the shockwave from the bounce of the compressed material, halted at nuclear densities, spreads out into a convective region with infalling matter. The timescales and energy deposited by neutrino radiation and hydrodynamic transport within the convective region determine the characteristics of the explosion and the amount of fall-back material accreted on the proto-NS at the core. This, along with structure of the star, determines the mass distribution of the resulting remnants. Additional considerations of the core-collapse engine, such as collapsar or magnetar formation, will also play a role.

Such astrophysical BHs are not expected to have masses below $\sim 2.5 M_{\odot}$.  Neutron star masses are subject to the Tolman--Oppenheimer--Volkoff (TOV) stability limit, with suitable nuclear equation state. Commonly discussed TOV limits are $\sim 2 M_{\odot}$ for nonrotating and $\sim 2.5 M_{\odot}$ for rotating NSs.  Stellar  collapse can produce a $ 1.5 - 2.5 M_{\odot}$ NS~(e.g.~\cite{Godzieba:2020tjn}).  Observations of low-mass X-ray binaries show that the smallest BHs have masses in the $\sim 5-10 M_{\odot}$ range, with no BHs lying in the (first) ``mass gap'' of $\sim 3 - 5 M_{\odot}$~\cite{Ozel:2010su}. BHs residing within the mass gap are expected from some stellar population synthesis models~(e.g.~\cite{Ertl:2019zks,Woosley:2020mze}) and can naturally appear from delayed core-collapse supernovae \cite{Fryer:2012}. Detection of compact remnants within the mass gap, whose origin can be distinguished through gravity wave signal rates as well as accompanying electromagnetic counterparts \cite{Drozda:2020qab}, could give insight into the supernova explosion mechanism \cite{Belczynski:2012}. Detection of BHs below $\sim 3 M_{\odot}$ would point to an origin different from conventional stellar evolution, possibly indicating new physics.

Solar-mass BHs can be of primordial origin (PBHs), directly produced in the early Universe, prior to the formation of galaxies and stars. Such PBHs could also contribute to dark matter (DM) (e.g.~\cite{Zeldovich:1967,Hawking:1971ei,Carr:1974nx,GarciaBellido:1996qt,Khlopov:2008qy,Frampton:2010sw,Kawasaki:2016pql,Cotner:2016cvr,Cotner:2017tir,Carr:2016drx,Inomata:2016rbd,Pi:2017gih,Inomata:2017okj,Garcia-Bellido:2017aan,Inoue:2017csr,Georg:2017mqk,Inomata:2017bwi,Kocsis:2017yty,Ando:2017veq,Cotner:2018vug,Sasaki:2018dmp,Carr:2018rid,Cotner:2019ykd,Bird:2016dcv,Kusenko:2020pcg,Lu:2020bmd}).  In particular, solar-mass BHs can readily appear in some models if they are associated with the mass inside the horizon at the QCD epoch~(e.g.~\cite{Dolgov:2013lba,Dolghov:2020hjk}). 

Solar-mass BHs can also be ``transmuted''~\cite{Takhistov:2017bpt}, naturally arising as a consequence of NS implosions\footnote{Even smaller solar-mass BHs can appear from implosions of white dwarfs. Search of sub-solar mass BHs has been recently performed at LIGO/VIRGO~\cite{LIGO:2019qbw}.} caused by capture of sublunar-mass PBHs  (e.g.~\cite{Fuller:2017uyd,Capela:2013yf,Takhistov:2017bpt,Bramante:2017ulk,Takhistov:2017nmt,Capela:2013yf,Genolini:2020ejw}). These events could be the origin of a variety of interesting astrophysical signals, like orphan kilonovae and GRBs or 511 keV emission~\cite{Fuller:2017uyd,Takhistov:2017nmt}. More importantly, the relevant small PBHs lie in the open parameter space window where PBHs can constitute all of the DM. Transmuted solar-mass BHs can also arise due to accumulation of particle DM and subsequent implosion of NSs. This can occur if the DM particle interacts with baryons and is nonannihilating~(e.g.~\cite{Bramante:2017ulk,Kouvaris:2018wnh}). These requisite properties can be achieved with asymmetric DM models \cite{NUSSINOV:1985a,Kaplan:2009ag,Petraki:2013wwa,Zurek:2013wia}, including those based on bosons~(e.g.~\cite{McDermott:2011jp,Bell:2013xk}) or self-interacting fermions~(e.g.~\cite{Kouvaris:2011gb,Bramante:2013nma}).~Here, the relic abundance is set by an initial particle-antiparticle asymmetry that could also be connected with the observed baryon asymmetry of the Universe. 

Given the variety of solar-mass BH production possibilities, and given the uncertainties surrounding the true nature of the $\sim 1.5-2.5 M_{\odot}$  solar-mass constituent events in mergers (e.g.~\cite{Takhistov:2017bpt}) and distinguishing BHs from compact stars, an obvious set of questions arises. What data are needed to establish or rule out the existence of solar-mass BHs with an origin distinct from conventional stellar evolution production routes? Do the currently extant data provide some insight into this issue? The recent GW compact object merger events have sparked research on this and related topics~\cite{Jedamzik:2020omx,Clesse:2020ghq,Tsai:2020hpi}.

\begin{figure*}[tb]
  \includegraphics[width=0.45\linewidth]{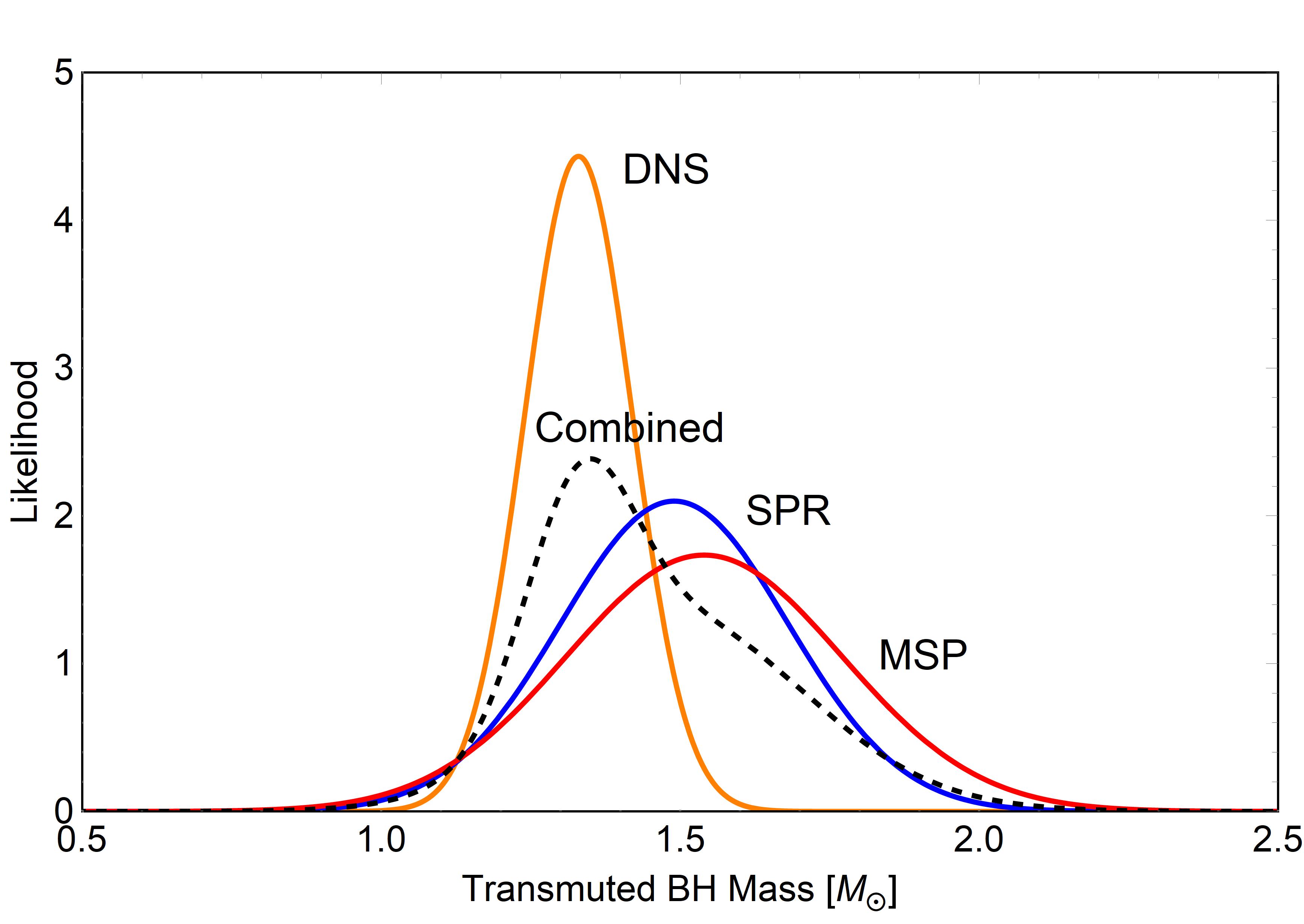}
  \hspace{1cm}
  \includegraphics[trim={0mm 0mm 0 0mm},clip,width=.45\textwidth]{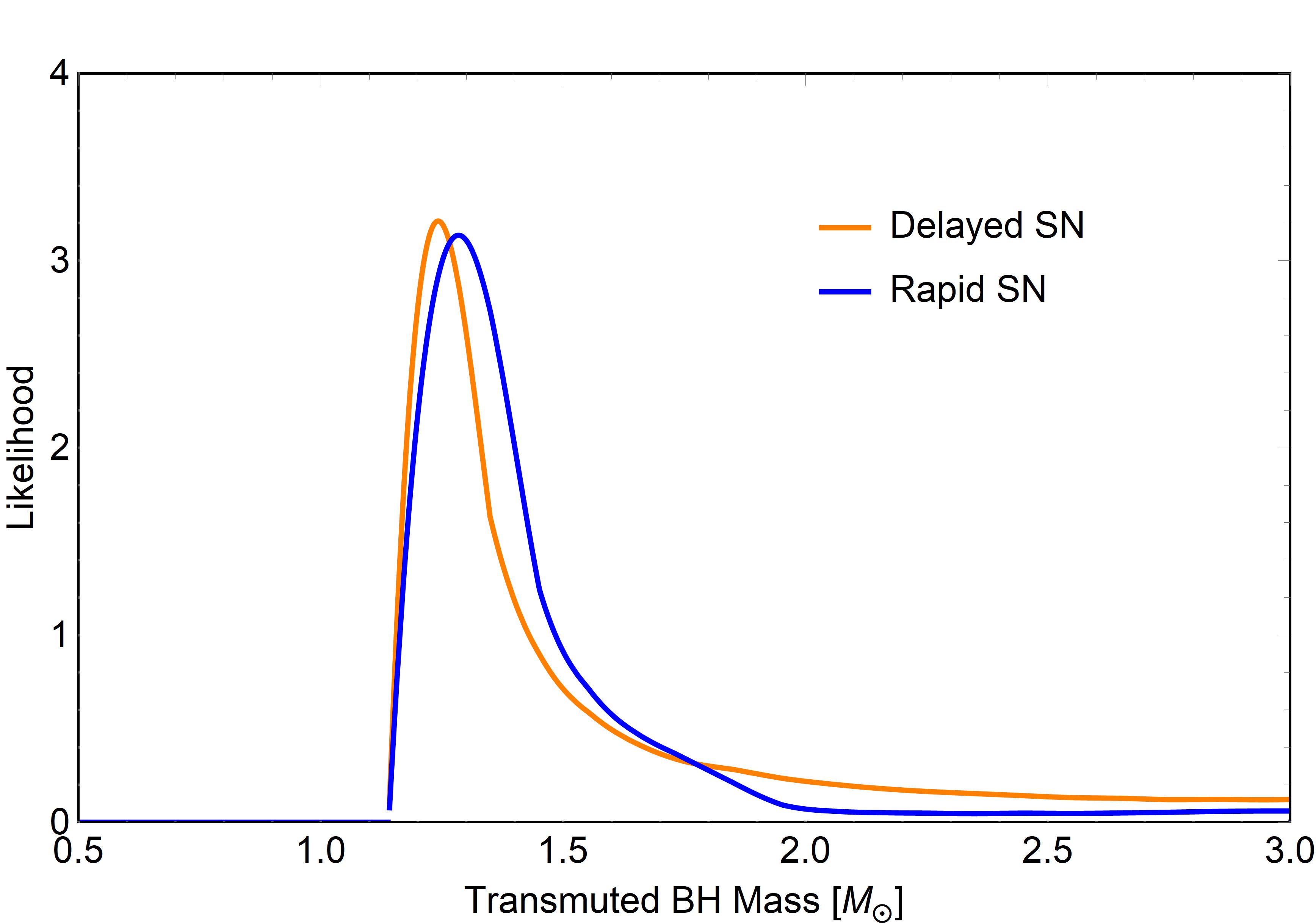}
\caption{Expected mass distribution of transmuted solar-mass BHs assuming that these track their NS progenitors. \textbf{Left:} Considered subpopulations of imploding NSs from slow pulsars (SPR, blue), recycled millisecond pulsars (MSP, red) and double neutron stars (DNS, orange) as well as combined distribution (black dashed) are shown. Input parameters for Gaussian distributions of the NS populations are taken from~Ref.~\cite{Kiziltan:2013oja,Ozel:2016oaf}. \textbf{Right:} Imploding NSs from models of delayed and rapid collapse supernovae \cite{Fryer:2012}, assuming solar metallicity of stellar progenitors.}
\label{fig:nsmass}
\end{figure*}

In this work, we establish a decisive test that could shed light on the origin of solar-mass BHs. As we point out, a unique feature of transmuted solar-mass BHs from particle or PBH DM interactions is that their mass distribution tracks the mass distribution of NSs. In contrast, while the mass distribution of solar-mass PBHs from the early Universe could be peaked or more broadly distributed, depending on the formation epoch and the details of the considered cosmological model, their mass distribution is unrelated to the stellar one.  

For transmuted solar-mass BHs, the capture rate of sublunar-mass PBHs\footnote{The base PBH capture rate is further enhanced when the NS velocity distribution is taken into account~\cite{Fuller:2017uyd}.} on a NS is given by~\cite{Capela:2013yf,Genolini:2020ejw}
\begin{align} \label{eq:pbhcap}
F_{\rm PBH} =&~ \sqrt{6 \pi}  \dfrac{\rho_{\rm DM}}{M_{\rm PBH} \overline{v}} \Big(\dfrac{2 G M_{\rm NS} R_{\rm NS}}{1 - 2 G M_{\rm NS}/R_{\rm NS}}\Big) \notag\\
&\times \Big(1- {\rm Exp}\Big[-\dfrac{3 E_{\rm loss}}{M_{\rm PBH}\overline{v}^2}\Big]\Big)~, 
\end{align}
where $M_{\rm NS}$ is the NS mass, $R_{\rm NS}$ is the NS radius, $M_{\rm PBH}$ is the mass of the PBH, $\overline{v}$ is the PBH velocity dispersion, $E_{\rm loss} \simeq 58.8 G^2 M_{\rm PBH}^2 M_{\rm NS}/R_{\rm NS}^2$ is the average interaction energy loss of a PBH in the NS and $\rho_{\rm DM}$ is the PBH contribution to the dark matter mass density. After capture, a PBH will settle within the NS and subsequently consume it. These processes would typically occur on significantly shorter timescales than the age of a system.

For the case of transmuted solar-mass BHs from particle DM accumulation, 
the rate with which DM particles are captured by a NS is given by \cite{Goldman:1989,Kouvaris:2007ay}
\begin{equation} \label{eq:pdmcap}
F_{\rm PDM} = \sqrt{6 \pi}  \dfrac{\rho_{\rm DM}}{m \overline{v}} \Big(\dfrac{2 G M_{\rm NS} R_{\rm NS}}{1 - 2 G M_{\rm NS}/R_{\rm NS}}\Big) \dfrac{\sigma}{\sigma_{\rm crit}}
\end{equation}
where $m$ is the DM particle mass, $\rho_{\rm DM}$ is the DM density,  $\overline{v}$ is the DM velocity dispersion, $R$ and $R_s$ the
NS radius and its Schwarzschild radius, respectively, and $(\sigma/\sigma_{\rm crit}) \leq 1$ denotes the efficiency factor for a given DM interaction cross section $\sigma$. Here $\sigma_{\rm crit} = 0.45 m_n R^2/M \simeq 10^{-45}$~cm$^{-2}$ is the critical
cross section above which on average every DM particle passing through the NS is captured. The captured DM thermalizes and collapses to a small BH if the number of particles grows beyond a critical value. Subsequently, the host star is rapidly consumed. As for the case of PBHs, these processes should also occur on significantly shorter timescales than the age of a system.

Transmuted solar-mass BHs should track the mass distribution of the NS population. The population of NSs is not uniform and can be classified into several categories. These include recycled fast-rotating millisecond pulsars (MSPs), slow pulsars (SPRs), and those in double NS binaries (DNSs). For the NS mass distribution, we adopt the established fit model described by a  Gaussian\footnote{The possibility of a bimodal mass distribution for MSPs
has also been suggested~\cite{Antoniadis:2016hxz,Ertl:2019zks,Woosley:2020mze}.} probability density function~\cite{Kiziltan:2013oja,Ozel:2016oaf}
\begin{equation} \label{eq:nspop}
    P (M_{\rm NS}) = \dfrac{1}{\sqrt{2 \pi \sigma_{\rm NS}^2}} {\rm Exp}\Big[-\dfrac{(M_{\rm NS} - M_0)^2}{2 \sigma_{\rm NS}^2}\Big]~,
\end{equation}
where the NS subpopulations are given by $M_0 = 1.33 M_{\odot}$ and $\sigma_{\rm NS} = 0.09 M_{\odot}$ for the DNSs, $M_0 = 1.54 M_{\odot}$ and $\sigma_{\rm NS} = 0.23 M_{\odot}$ for the MSPs, and $M_0 = 1.49 M_{\odot}$ and $\sigma_{\rm NS} = 0.19 M_{\odot}$ for SPRs. Using the NS population mass distributions, and assuming that these are tracked by transmuted solar-mass BHs, we show the expected transmuted BH mass distributions in Fig.~\ref{fig:nsmass}.  For the normalized combined distribution we have assumed equal contributions from each subpopulation. In addition to the basic Gaussian model, we also consider (see Fig.~\ref{fig:nsmass}) compact remnant mass distribution that is expected from theoretical models of delayed and rapid collapse supernovae that can fit the observed data \cite{Fryer:2012}. Solar metallicity is assumed for stellar progenitors. In these models, distribution extends to higher masses and has a steep cutoff at lower mass. A dip in the distribution is observed in rapid SN models around $\sim 2 M_{\odot}$.

 \begin{table*}[t]
\centering
\begin{tabular}{c|c|c|c|c|c|c} 
 \hline
  \hline
  Trial fit    & Candidate & Total events  & BH mass    & Gaussian K-S test & Delayed SN K-S test & Rapid SN K-S test \\   
        & GW events &  &  ($M_{\odot}$)   &   (p-value) &  (p-value) &  (p-value) \\ 
 \hline
  \hline
  \multirow{2}{*}{1}   & GW190814 & \multirow{2}{*}{2} & 2.6 & \multirow{2}{*}{$7.2 \times 10^{-3}$}  & \multirow{2}{*}{$7.9 \times 10^{-2}$} & \multirow{2}{*}{$1.5 \times 10^{-2}$} \\
     & GW190425 & & 1.8 &  & &   \\
    \hline
   & GW190814 & & 2.6 &  & &   \\
  2  & GW190425 & 3 & 1.8 &  $9.5 \times 10^{-2}$  & $7.4 \times 10^{-2}$ & $3.1 \times 10^{-2}$ \\
    & GW170817 & & 1.5 &  & &   \\ 
        \hline
     \multirow{7}{*}{3} & test event 1 & \multirow{7}{*}{7} & 1.3 &  \multirow{7}{*}{$1.8 \times 10^{-3}$} &  \multirow{7}{*}{$2.6 \times 10^{-2}$} &  \multirow{7}{*}{$3.4 \times 10^{-3}$}  \\
     & test event 2 &   & 1.5 &  & &   \\ 
     & test event 3 &   & 1.8 &  & &    \\ 
     & test event 4 &   & 2.0 &  & &    \\ 
     & test event 5 &   & 2.1 &  & &    \\ 
     & test event 6 &   & 2.2 &  & &    \\ 
     & test event 7 &   & 2.3 &  & &    \\ 
 \hline
  \hline
\end{tabular}
\caption{Fit of transmuted solar-mass BH candidate events to a mass distribution following the NS population.}
\label{tab:bhfit}
\end{table*}

In principle, the expected observable mass distribution of transmuted solar-mass BHs could be altered from that of NSs by the NS mass dependence of the DM capture rate or the GW detection sensitivity. From Eq.~\eqref{eq:pbhcap} and Eq.~\eqref{eq:pdmcap} it can be seen that the DM capture rate does not have a strong dependence on NS mass and radius variation in the range of interest, being altered by a factor of few\footnote{NS implosion details, such as asymmetry during PBH NS evolution \cite{Genolini:2020ejw}, could also potentially play a role on the transmuted BH mass distribution. However, since we do not expect significant mass emission during implosion compared to the total NS mass, such effects are not expected to drastically alter our conclusions.}.
For GWs from a binary merger, assuming a detection signal-to-noise ratio of 8, the inspiral signal horizon distance is given by~\cite{LIGO:2012aa}
\begin{equation} 
\mathcal{D} = \dfrac{1}{8}\Big(\dfrac{5 \pi}{24}\Big)^{1/2}\Big(G \mathcal{M}\Big)^{5/6} \pi^{-7/6} \sqrt{4 \int_{f_{\rm low}}^{f_{\rm high}} \dfrac{f^{-7/3}}{S_n(f)}df}~,
\end{equation}
where  $\mathcal{M} = [q/(1+q)^2]^{3/5} M$ is the chirp mass for a binary of total mass $M = m_1 + m_2$ and mass ratio $q = m_1/m_2$, $f$ is the GW frequency, $S_n(f)$ is the power spectral density denoting mean noise fluctuations at a given frequency. The upper limit on the frequency is set by the innermost circular orbit $f_{\rm ISCO} = 1/6\sqrt{6 \pi} G M$, while the lower is set by the detector threshold. Consequently, the GW detection sensitivity is not strongly affected by variation in NS mass. Hence, the mass distribution of transmuted solar-mass BHs is expected to track NS mass distribution without significant deviations. 

The utility of the proposed test can be demonstrated by fitting the solar-mass BH candidate events to the NS mass distribution, for which we employ a Kolmogorov-Smirnov (K-S) test. For the null hypothesis we assume that data follows the combined NS population distribution, including the PSR, MSP and DNS subpopulations described above, as displayed in Fig.~\ref{fig:nsmass}. We first test the Gaussian compact remnant model (see trial fit 1 in Tab.~\ref{tab:bhfit}) with the hypothesis that the recently observed $\sim 2.6 M_{\odot}$ GW190814 (BH-BH?) and $\sim 1.8 M_{\odot}$ GW190425 (BH-BH?) solar-mass BH candidate GW events follow the NS mass distribution. The resulting K-S p-value of $7.2 \times 10^{-3}$ strongly rejects the null hypothesis at the $>99\%$ level. 

For the next trial test, trial fit 2 in Tab.~\ref{tab:bhfit}, we consider
the GW190814 and GW190425 events together with an additional $\sim 1.5 M_{\odot}$ event assuming that one of the NSs in the GW170817 NS-NS merger event is actually a solar-mass BH, a possibility still allowed by observations. The newly obtained p-value of $9.5 \times 10^{-2}$ weakens the rejection of the null hypothesis to $\sim 90\%$, but still remains significant. 

Finally, in trial fit 3 in Tab.~\ref{tab:bhfit}, we display the test result assuming a sample population of solar-mass BHs with several events with mass above $1.5 M_{\odot}$, resulting in a $\mathcal{O}(10^{-3})$ p-value. These considerations highlight the importance of detecting larger $>1.5 M_{\odot}$ solar-mass BH candidate events. Moreover, they also highlight the inconsistency of the $\sim 2.6 M_{\odot}$ candidate event as revealed by testing with the NS mass distribution.

Assuming a rapid or delayed supernova model for compact remnants, the same general trend is observed. However, discrimination is weaker for delayed SN models stemming from the  sizable tail extending to larger masses. It is known that delayed SN models can populate the ``mass gap'' \cite{Fryer:2012,Belczynski:2012}.  This signifies the importance of understanding theoretically standard astrophysical formation mechanisms of compact remnants. Thus, while there are sizable uncertainties on the masses of detected binary constituents, we have demonstrated that the observed solar-mass BH candidate events are unlikely to be from NS DM implosions.  

The mass function of PBHs depends on the production scenario. Some models predict a broad distribution of masses, such as a piecewise power law distributions~\cite{Deng:2016vzb,Deng:2017uwc,Kusenko:2020pcg}, which can be easily distinguished from the expected mass function of transmuted black holes.  Some other models predict a narrow mass function~\cite{Frampton:2010sw,Kawasaki:2012kn,Cotner:2016cvr,Cotner:2016dhw,Cotner:2017tir,Cotner:2018vug,Cotner:2019ykd}, which can peak at any point over a broad range of masses.  However, it is extremely unlikely that the maximum of such a narrow mass distribution could coincide with the predicted peak of the transmuted mass function. In the unlikely scenario that the peaks of distributions coincide, with sufficient statistics, our test will still be able to discriminate between them based on the shape of the distribution.

More sophisticated analyses and inclusion of possible uncertainties will provide a further improvement of our proposed test. The test will become increasingly powerful with upcoming observations, as higher statistics for binary events will allow differentiation between BHs and NSs with significant confidence. This will be true even when only $<100$~events are detected~\cite{Chen:2019aiw}. Additional discriminating factors, such as coincidence signatures, event rates and spatial distribution~(see e.g.~\cite{Fuller:2017uyd,Takhistov:2017bpt,Bramante:2017ulk,Takhistov:2017nmt}), could be employed to strengthen the efficacy of this proposed test.

In conclusion, we showed that transmuted solar-mass BHs from NS implosions, caused by capture of sublunar-mass PBHs or an accumulation of particle DM, can be identified because they track the NS mass distribution. This is different from the typical predictions and expectations of primordial solar-mass BHs created in the early Universe. Several recent candidate events have been identified. We have presented mass distribution analysis of solar-mass BHs as a powerful novel test of their origin. As we have shown, it is unlikely that the recently detected events originate from NS implosions. This statistical test method provides a new tool for analyzing ideas and models for DM and related solar-mass BHs.
Our study highlights the importance of further progress in theoretical understanding of standard astrophysical formation mechanisms of compact remnants.

We thank David Radice for discussions.
The work of V.T. and A.K. was supported by the U.S. Department of Energy (DOE) Grant No.~DE-SC0009937. V.T. and A.K. were also supported by the World Premier International Research Center Initiative (WPI), MEXT, Japan. A.K. was also supported by Japan Society for the Promotion of Science (JSPS) KAKENHI Grant No. JP20H05853. G.M.F. acknowledges support from NSF Grant No. PHY-1914242, the N3AS Physics Frontier Center, NSF Grant No. PHY-2020275 and the Heising-Simons Foundation (2017-228), and from the U. S. Department of Energy Scientific Discovery through Advanced Computing (SciDAC-4) grant register No. SN60152 (award number de-sc0018297).

 
\bibliography{bibliography}
 
\end{document}